\documentclass[10pt]{article}
\usepackage[OE]{express}
\usepackage{subfigure}

\begin{document}
\title{Generation of long-living entanglement between two distant
three-level atoms in non-Markovian environments}

\author{Chuang Li,\authormark{1} Sen Yang,\authormark{1} Yan Xia,\authormark{2} Jie Song,\authormark{1,3} and Wei-Qiang Ding\authormark{1,4}}

\address{\authormark{1}Department of Physics, Harbin Institute of Technology, Harbin, 150001, China\\
\authormark{2}Department of Physics, Fuzhou University, Fuzhou, 350002, China\\
\authormark{3}jsong@hit.edu.cn\\
\authormark{4}wqding@hit.edu.cn}



\begin{abstract}
In this paper, a scheme for the generation of long-living entanglement
between two distant $\Lambda$-type three-level atoms separately trapped in
two dissipative cavities is proposed. In this scheme, two dissipative
cavities are coupled to their own non-Markovian environments and two
three-level atoms are driven by the classical fields. The entangled state
between the two atoms is produced by performing Bell state measurement (BSM)
on photons leaving the dissipative cavities. Using the time-dependent
Sch\"{o}rdinger equation, we obtain the analytical results for the evolution
of the entanglement. It is revealed that, by manipulating
the detunings of classical field, the long-living stationary entanglement
between two atoms can be generated in the presence of dissipation.
\end{abstract}



\section{Introduction}
Quantum entanglement, as the most important resource for quantum science and
technology, draws a great deal of attention in various domains \cite{1}, such
as quantum teleportation \cite{2}, quantum dense coding \cite{3}, quantum
cryptography \cite{4}, and quantum computation \cite{5}. Therefore, many
schemes have been proposed to generate entangled states, such as
trapped ions \cite{6,7}, quantum electrodynamics \cite{8,9}, and photon
pairs \cite{10,11}.

In order to complete a quantum operation, the long-living entanglement is
needed. In real physical systems, however, quantum entanglement is fragile
and very easy to be destroyed due to the interaction between quantum system
and environments \cite{12,13,14}. Therefore, many efforts have been devoted to
the dynamical evolution of entanglement in Markovian environments \cite{15,16,17,18}.
In contrast, non-Markovian dynamics shows more interesting
phenomena because of the memory effect, and has been used in various quantum
operations \cite{18a,19,20,20a,21}. Up to now, extensive researches on the entangled states
for two-level atoms in dissipative environments have been done \cite{22,23,24,25}.
For example, in \cite{25}, Nourmandipour et al. investigate the entanglement
swapping between two two-level atoms. Their results show that the
stationary entanglement between two two-level atoms can be generated in the
presence of dissipation.

Compared with the two-dimensional entanglement, high-dimensional entangled states
are more competitive due to the fact that three-level quantum systems provide more
secure quantum key distributions than those based on two-level systems \cite{26,27,28}.
Therefore, extensive researches have been devoted to the generation of three-dimensional
entanglement \cite{29,30,31,32,33}. For example, in \cite{32}, the generation of
three-dimensional entanglement of two distant atoms in Markovian environments is proposed.
In practice, the dissipation of cavities is unavoidable, and generating three-dimensional entangled
states in non-Markovian environments is valuable and worth studying.

In this paper, we propose a scheme for producing the entanglement between two atoms separately
trapped in two dissipative cavities. We first investigate the dynamical evolution of a
three-level atom in non-Markovian environments by using the time-dependent Sch\"{o}rdinger
equation. Then, we generate the entanglement between two atoms by performing Bell state
measurement on photons leaving the cavities. We use negativity to quantify the amount of
entanglement \cite{34} and discuss the effect of detunings and initial atomic states on the
evolution of entanglement. The rest of the paper is organized as follows:
In Sec. \uppercase\expandafter{\romannumeral2},
we introduce the model of the atom-field coupling system, and the dynamical evolution of
entanglement between the atom and the cavity field is presented in
Sec. \uppercase\expandafter{\romannumeral3}.
In Sec. \uppercase\expandafter{\romannumeral4},
we produce the entangled state between two atoms
by performing Bell state measurement and discuss
the effect of detunings and initial atomic states on the evolution
of entanglement. The conclusions are drawn in Sec. \uppercase\expandafter{\romannumeral5}.

\section{THE MODEL}

\begin{figure}[htbp]
  \centering
\includegraphics[width=0.5\textwidth]{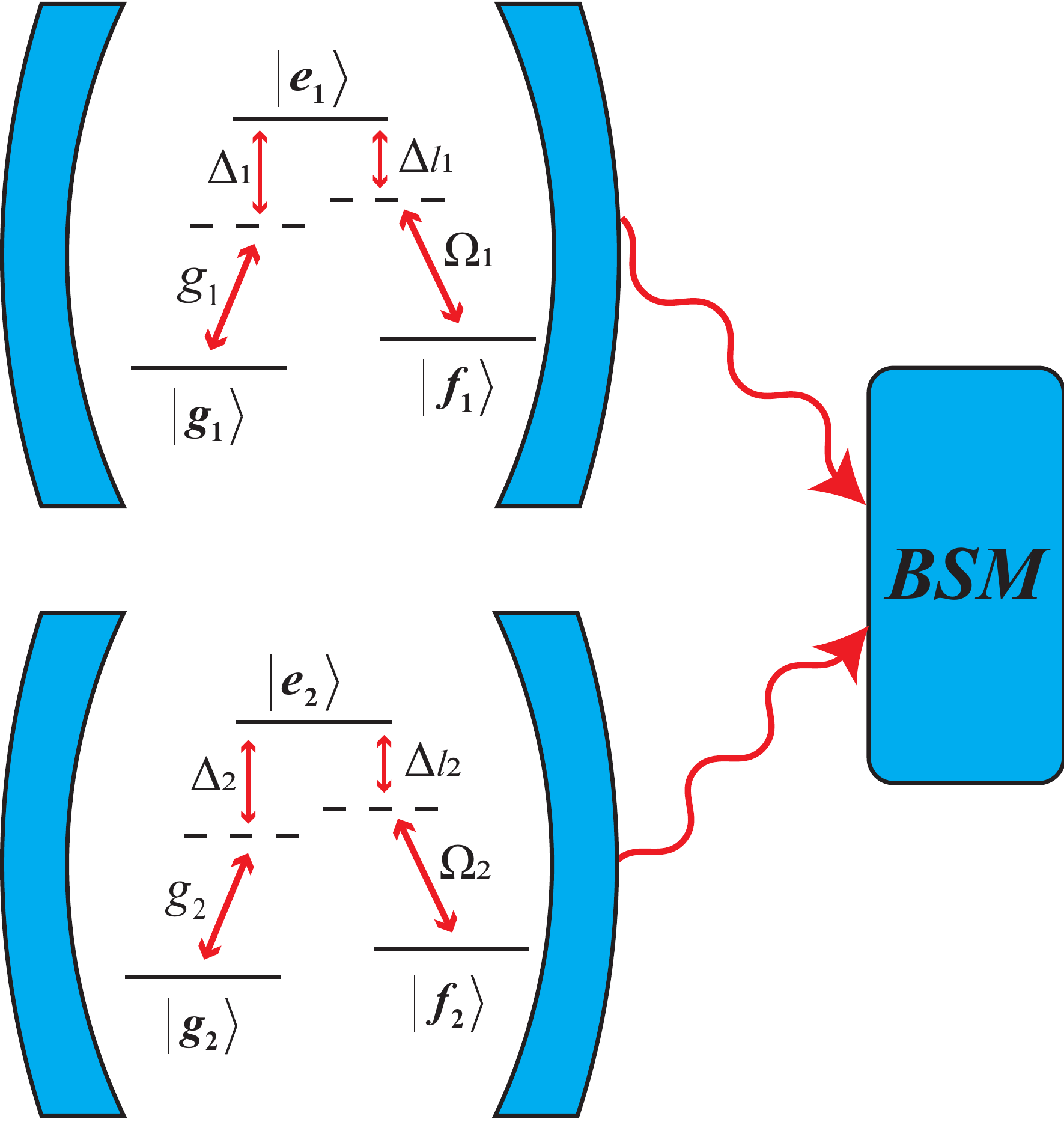}
 \caption{Schematic representation of the setup. BSM is performing Bell
 state measurement on photons leaving the cavities.\label{fl1}}
\end{figure}

We consider a system formed by two separate dissipative cavities,
each of which contains a $\Lambda$-type three-level atom with ground state
($|g\rangle$), lower and upper excited states ($|f\rangle$,
$|e\rangle$) (see Fig. \ref{fl1}). The quantum states $|g_{i}\rangle$,
$|f_{i}\rangle$, and $|e_{i}\rangle$ ($i=1,2$) have the energies of $\omega_{g_{i}}$,
$\omega_{f_{i}}$, and $\omega_{e_{i}}$, respectively ($\hbar=1$). We assume both two dissipative cavities have high quality factors. In
the $i$th cavity, the transition $|g_{i}\rangle\leftrightarrow|e_{i}\rangle$
is coupled to a single-mode cavity field with the coupling constant
$g_{i}$, while the transition $|f_{i}\rangle\leftrightarrow|e_{i}\rangle$
is driven through a classical field with the coupling constant $\Omega_{i}$.
Assuming that the cavity field interacts with a reservoir consisting of
a set of continuous harmonic oscillators, the Hamiltonian describing the
field-reservoir is given by
\begin{equation}\label{1}
H_{c_{i}}=\omega_{c_{i}} a_{i}^{\dag}a_{i}+\int_0^\infty B_{i}^{\dag}(\eta)B_{i}(\eta)\mathrm{d}\eta+\int_0^\infty G_{i}(\eta)[a_{i}^{\dag}B_{i}(\eta)+\mathrm{H.c.}]\mathrm{d}\eta,
\end{equation}
where $\omega_{c_{i}}$ is the frequency of the cavity field,
$G_{i}(\eta)$ is the coupling strength between the cavity field and the
reservoir, which is a function of frequency $\eta$. $B_{i}^{\dag}(\eta)$
(and $B_{i}(\eta)$) is the creation (and annihilation) operator
of the reservoir, which obeys the commutation relation of
$[B_{i}(\eta),B_{j}^{\dag}(\eta^{'})]=\delta_{ij}\delta(\eta-\eta^{'})$. The model of the field-reservoir shows that the dissipative cavity has a Lorentzian spectral density implying the nonperfect reflectivity of the cavity mirrors.
Supposing that the reservoir has a narrow bandwidth, we can
extend integrals over $\eta$ from $0$ to $-\infty$ and take $G_{i}(\eta)$ as
a constant. Thus, by introducing the dressed operator
$A_{i}(\omega)=\alpha_{i}(\omega)a_{i}+\int_{-\infty}^\infty\beta_{i}(\omega,\eta)B_{i}(\eta)\mathrm{d}\eta$,
one is able to diagonalize the Hamiltonian (\ref{1}) as \cite{35}
\begin{equation}
H_{c_{i}}=\int_{-\infty}^\infty\omega A^{\dag}_{i}(\omega)A_{i}(\omega)\mathrm{d}\omega.
\end{equation}
The annihilation operator $a_{i}$ is given by
\begin{equation}
a_{i}=\int_{-\infty}^\infty\alpha_{i}^{*}(\omega)A_{i}(\omega)\mathrm{d}\omega,
\end{equation}
with
\begin{equation}
\alpha_{i}(\omega)=\frac{\sqrt{\kappa_{i}/\pi}}{\omega-\omega_{c_{i}}+i\kappa_{i}},
\end{equation}
where $\kappa_{i}$ is the decay rate of the $i$th cavity. Consequently, the total Hamiltonian of the atom-field system is
\begin{equation}
\begin{aligned}
H_{i}=&\int_{-\infty}^\infty\omega A^{\dag}_{i}(\omega)A_{i}(\omega)\mathrm{d}\omega+\omega_{e_{i}}|e_{i}\rangle\langle e_{i}|+\omega_{f_{i}}|f_{i}\rangle\langle f_{i}|+\omega_{g_{i}}|g_{i}\rangle\langle g_{i}|\\
&+\int_{-\infty}^\infty g_{i}[\alpha^{*}_{i}(\omega)A_{i}(\omega)|e_{i}\rangle\langle g_{i}|+\mathrm{H.c.}]\mathrm{d}\omega+\Omega_{i}[|e_{i}\rangle\langle f_{i}|e^{-i\omega_{l_{i}}t}+\mathrm{H.c.}],
\end{aligned}
\end{equation}
where $\omega_{l_{i}}$ is the frequency of the classical field in the $i$th cavity. Without loss of generality,
we assume the atoms and the cavities have the same parameters, i.e.,
$\omega_{e_{1}}=\omega_{e_{2}}\equiv\omega_{e}$,
$\omega_{f_{1}}=\omega_{f_{2}}\equiv\omega_{f}$,
$\omega_{g_{1}}=\omega_{g_{2}}\equiv\omega_{g}$,
$\omega_{c_{1}}=\omega_{c_{2}}\equiv\omega_{c}$,
$\kappa_{1}=\kappa_{2}\equiv\kappa$,
$\omega_{l_{1}}=\omega_{l_{2}}\equiv\omega_{l}$,
$g_{1}=g_{2}\equiv g$, and
$\Omega_{1}=\Omega_{2}\equiv\Omega$.
In the interaction picture, the interaction Hamiltonian is given by
\begin{equation}
H_{I_{i}}=\int_{-\infty}^\infty g[\alpha^{*}(\omega)A(\omega)|e_{i}\rangle\langle g_{i}|e^{i(\omega_{e}-\omega_{g}-\omega)t}+\mathrm{H.c.}]\mathrm{d}\omega+\Omega[|e_{i}\rangle\langle f_{i}|e^{-i\Delta_{l}t}+\mathrm{H.c.}],
\end{equation}
where $ \Delta_{l}=\omega_{l}-(\omega_{e}-\omega_{f}) $ is the detuning
of the classical field.
Assuming the atom is initially in the coherent superposition of the
quantum states $|f_{i}\rangle$ and $|g_{i}\rangle$, and the cavity field
is in the vacuum state $|0\rangle$, the initial wave function of
the subsystem is given by
\begin{equation}
|\psi(0)\rangle_{i}=[\cos(\theta_{i}/2)|f_{i}\rangle+\sin(\theta_{i}/2)e^{i\varphi_{i}}|g_{i}\rangle]|0\rangle_i,
\end{equation}
where $\theta_{i}\in[0,\pi]$, $\varphi_{i}\in[0,2\pi]$ and $|0\rangle_i$ represents for the vacuum state of the $i$ environments. $|1_\omega\rangle_i=A^{\dag}(\omega)|0\rangle_i$ represents that there is one photon at frequency $\omega$ in the $i$ environments.
With at most only one excitation, the wave function of the subsystem
at any time $t$ can be written as
\begin{equation}\label{2}
|\psi(t)\rangle_{i}=[E_{i}(t)|e_{i}\rangle+F_{i}(t)|f_{i}\rangle+G_{i}(t)|g_{i}\rangle]|0\rangle_i+\int_{-\infty}^\infty U_{i}(t,\omega)|g_{i}\rangle|1_\omega\rangle_i\mathrm{d}\omega,
\end{equation}
where $E_{i}(t)$, $F_{i}(t)$, $G_{i}(t)$, and $U_{i}(t)$ are the probability
amplitudes which should be determined. Using the Sch\"{o}rdinger
equation, we obtain
\begin{equation}\label{3}
\dot{E_{i}}(t)=-ig\int_{-\infty}^\infty\alpha^{*}(\omega)e^{i(\omega_{e}-\omega_{g}-\omega)t}U_{i}(\omega,t)\mathrm{d}\omega-i\Omega e^{-i\Delta_{l}t}F_{i}(t)
\end{equation}
\begin{equation}\label{4}
\dot{F_{i}}(t)=-i\Omega e^{i\Delta_{l}t}E_{i}(t)
\end{equation}
\begin{equation}\label{5}
\dot{G_{i}}(t)=0
\end{equation}
\begin{equation}\label{6}
\dot{U_{i}}(t)=-ig\alpha(\omega)e^{-i(\omega_{e}-\omega_{g}-\omega)t}E_{i}(t)
\end{equation}
The differential equations can be solved as
$G_{i}(t)=G_{i}(0)=\sin(\theta_{i}/2)e^{i\varphi_{i}}$. Performing
time integration of Eq. (\ref{4}) and Eq. (\ref{6}) and substituting
the results into Eq. (\ref{3}), we obtain
\begin{equation}\label{7}
\dot{E_{i}}(t)=-\int_{0}^{t}f(t-t_{1})E_{i}(t_{1})\mathrm{d}t_{1}-\Omega^{2}\int_{0}^{t}e^{-i\Delta_{l}(t-t_{2})}E_{i}(t_{2})\mathrm{d}t_{2}-i\Omega F_{i}(0)e^{-i\Delta_{l}t},
\end{equation}
where the correlation function $f(t-t_{1})=\int_{-\infty}^\infty J(\omega)e^{i(\omega_{e}-\omega_{g}-\omega)(t-t_{1})}\mathrm{d}\omega$.
$J(\omega)$ is the spectral densities, which is chosen as Lorentzian
function
\begin{equation}\label{7l}
J(\omega)=g^2|\alpha(\omega)|^{2}=\frac{1}{\pi} \frac{g^2\kappa}{(\omega-\omega_{c})^{2}+\kappa^{2}}.
\end{equation}
In Eq. (\ref{7l}), $\tau_g=g^{-1}$ is related to the relaxation time of the system and $\tau_{\kappa}=\kappa^{-1}$ is the correlation time of the reservoir. When the correlation time of the reservoir is greater than the relaxation time ($\tau_\kappa\gg\tau_g$), the system is coupled to non-Markovian environments. Conversely, when the relaxation time is greater than the correlation time of the reservoir ($\tau_g\gg\tau_{\kappa}$), the system is coupled to Markovian environments.
Substituting the spectral densities into the correlation function, the correlation
function can be written as
\begin{equation}
f(t-t_{1})=g^2e^{-(\kappa+i\Delta)(t-t_{1})},
\end{equation}
where $ \Delta=\omega_{c}-\omega_{e}+\omega_{g} $ is the detuning
of the cavity field. The integro-differential equation
(\ref{7}) can be written as
\begin{equation}\label{8}
\dot{E_{i}}(t)=-g^{2}\int_{0}^{t}e^{-(\kappa+i\Delta)(t-t_{1})}E_{i}(t_{1})\mathrm{d}t_{1}-\Omega^{2}\int_{0}^{t}e^{-i\Delta_{l}(t-t_{2})}E_{i}(t_{2})\mathrm{d}t_{2}-i\Omega F_{i}(0)e^{-i\Delta_{l}t}.
\end{equation}
In Eq. (\ref{8}), the first term is the interaction between the atom and the cavity
field, which leads to the dissipation of the quantum system; the
remaining terms are the interaction between the atom and the classical
field. With the help of Laplace transform, we solve the
integro-differential equation Eq. (\ref{8}) exactly. The result is expressed by
\begin{equation}
E_{i}(t)=F_{i}(0)\sum\nolimits_{k=1}^{k=3}c_{k}e^{s_{k}t},
\end{equation}
where $s_{k}$ is the $k$th root of the cubic equation
$s^{3}+[i(\Delta+\Delta_{l})+\kappa]s^{2}+[g^{2}+\Omega^{2}+i\Delta_{l}(i\Delta+\kappa)]s+\Omega^{2}(i\Delta+\kappa)+ig^{2}\Delta_{l}=0$,
and
$c_{1}=-i\Omega(s_{1}+i\Delta+\kappa)/((s_{1}-s_{2})(s_{1}-s_{3}))$,
$c_{2}=-i\Omega(s_{2}+i\Delta+\kappa)/((s_{2}-s_{1})(s_{2}-s_{3}))$,
and
$c_{3}=-i\Omega(s_{3}+i\Delta+\kappa)/((s_{3}-s_{2})(s_{3}-s_{1}))$.

\section{THE ENTANGLEMENT BETWEEN THE ATOM AND THE CAVITY FIELD}

Let us introduce the linear entropy to quantify the entanglement between the atom and the
cavity field, which is defined as
\begin{equation}
S_{A}(\theta,\varphi,t)=1-Tr(\rho_{A}^{2}),
\end{equation}
where $\rho_{A}$ is the atomic reduced density matrix of each subsystem.
The range of the linear entropy is between $0$ for pure state and $1-1/d$ for
completely mixed state, where $d$ is the dimension of the density matrix
(here $d=3$). Using Eq. (\ref{2}), we obtain the atomic reduced
density matrix $\rho_{A}$ as follows
\begin{equation}
\rho_{A}=
\begin{pmatrix}
|E_{i}(t)|^{2}&E_{i}(t)F_{i}^{*}(t)&E_{i}(t)G_{i}^{*}(t)\\
F_{i}(t)E_{i}^{*}(t)&|F_{i}(t)|^{2}&F_{i}(t)G_{i}^{*}(t)\\
G_{i}(t)E_{i}^{*}(t)&G_{i}(t)F_{i}^{*}(t)&1-|E_{i}(t)|^{2}-|F_{i}(t)|^{2}\
\end{pmatrix}.
\end{equation}
In this paper, we calculate the average
linear entropy with respect to all possible input states on the surface of
the Bloch sphere as\cite{36}
\begin{equation}
S_{A}^{av}(t)=\frac{1}{4\pi}\int S_{A}(\theta,\varphi,t)\sin(\theta)\mathrm{d}\theta\mathrm{d}\varphi.
\end{equation}

\begin{figure}[hthp]
\centering
\includegraphics[width=0.5\textwidth]{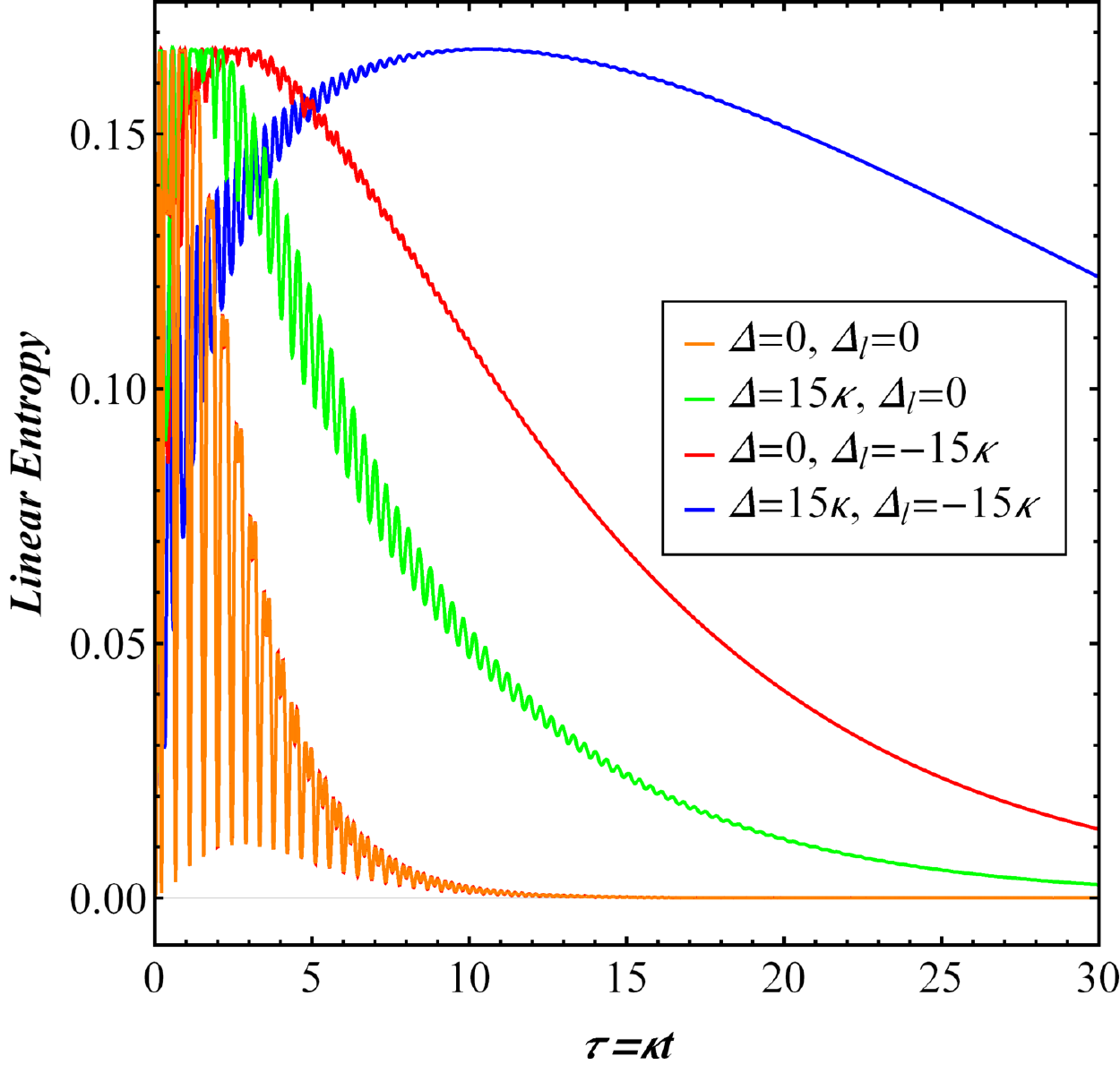}
\caption{ The evolution of the average linear entropy as a function of the scaled
 time $\tau=\kappa t$ for different detunings in non-Markovian environments:
 $\Delta=0, \Delta_{l}=0$ (orange curve), $\Delta=15\kappa, \Delta_{l}=0$ (green curve),
 $\Delta=0, \Delta_{l}=-15\kappa$ (red curve), and
 $\Delta=15\kappa, \Delta_{l}=-15\kappa$ (blue curve).
 Other parameters: $ g=\Omega=10\kappa$.\label{fl2}}
\end{figure}

\begin{figure}[htbp]
\centering
\subfigure[]{\includegraphics[width=0.4\textwidth]{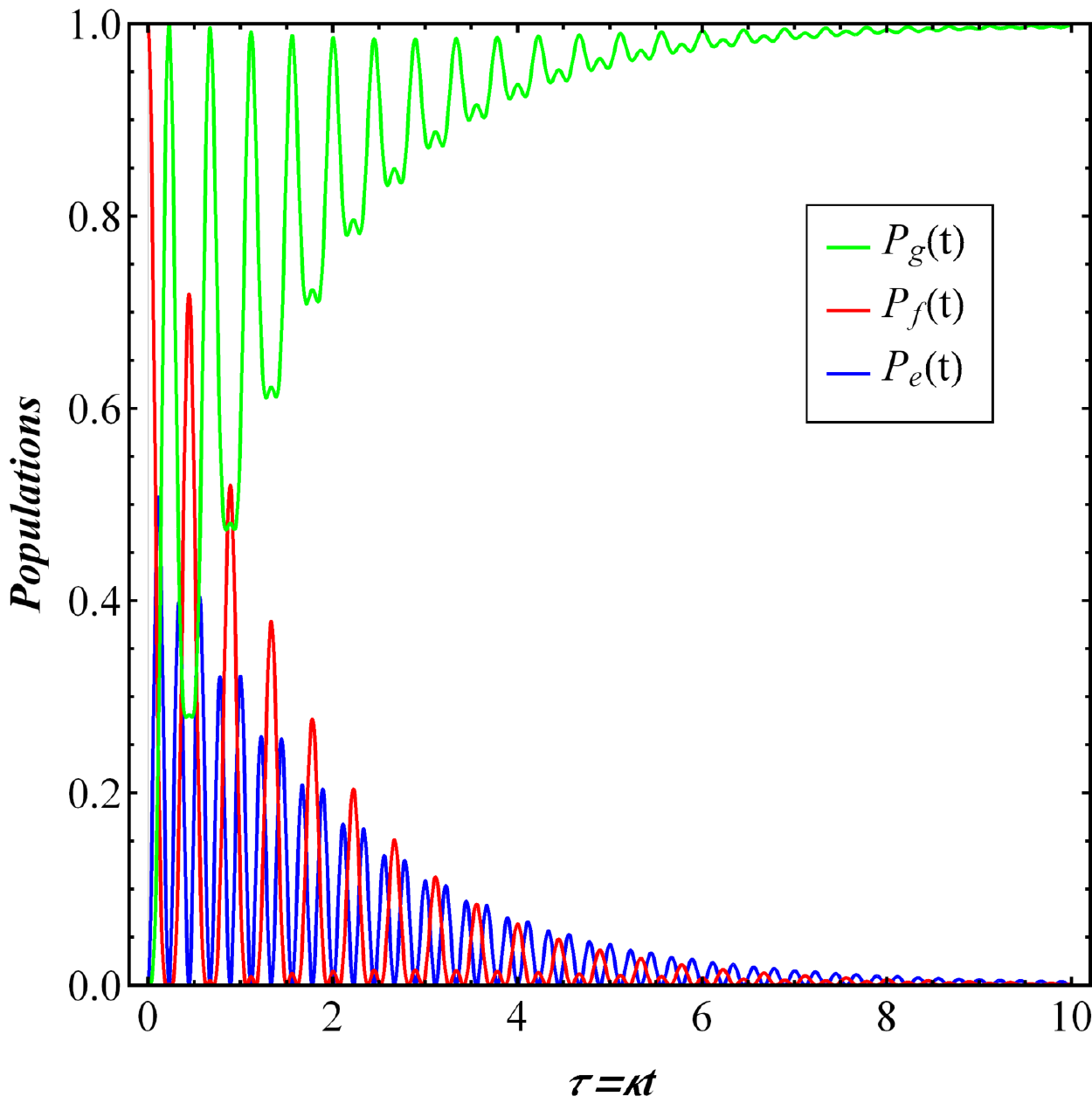}}
\hspace{2mm}
\subfigure[]{\includegraphics[width=0.4\textwidth]{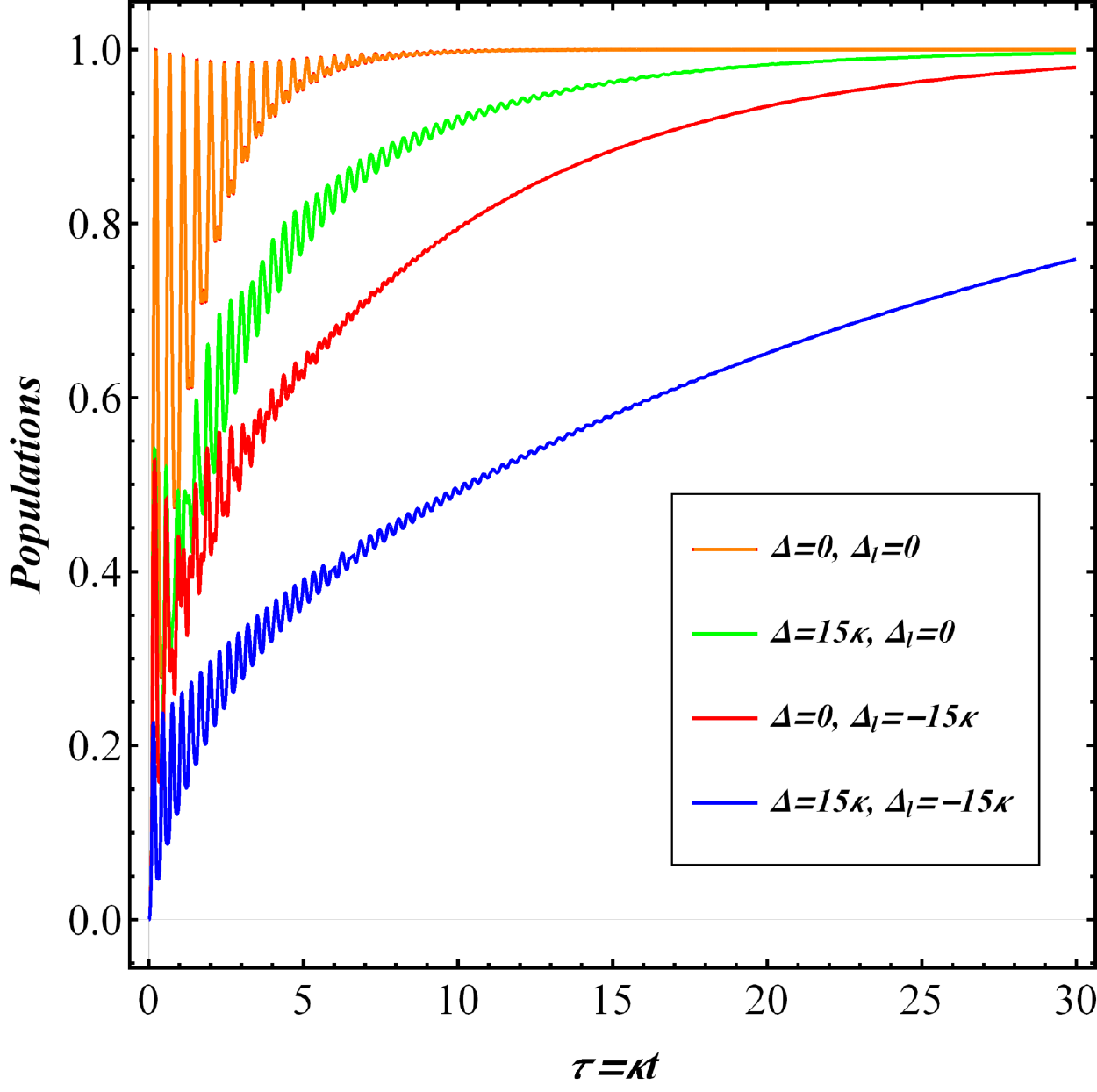}}
\caption{(a) The evolution of the populations of the states $|e\rangle$, $|f\rangle$, and $|g\rangle$ as a function of the scaled time $ \tau=\kappa t $ for the initial state $|f\rangle$ :
the population of the state $|e\rangle$ (blue curve),
the population of the state $|f\rangle$ (red curve), and
the population of the state $|e\rangle$ (green curve).
(b) The evolution of the populations of the state $|g\rangle$ as a function of the
 scaled time $ \tau=\kappa t $ for different detunings:
 $\Delta=0, \Delta_{l}=0$ (orange curve), $\Delta=15\kappa, \Delta_{l}=0$ (green curve),
 $\Delta=0, \Delta_{l}=-15\kappa$ (red curve), and
 $\Delta=15\kappa, \Delta_{l}=-15\kappa$ (blue curve).
Other common parameters: $\Delta=\Delta_{l}=0$, $g=\Omega=10\kappa$, and $\theta=\varphi=0$.\label{fl2b1}}
\end{figure}

Fig. \ref{fl2} illustrates the evolution of entanglement between the
atom and the cavity field over the scaled time $\tau=\kappa t$ in
non-Markovian environments. In Fig. \ref{fl2},
the average linear entropy exhibits an oscillatory behaviour for the
memory effect of non-Markovian environments. In the absence of the
detuning, the average linear entropy decays rapidly. In the
presence of the detunings, the average linear
entropy first increases to a maximum and then gradually decreases.
Fig. \ref{fl2b1}(a) shows the evolution of the populations of the states $|e\rangle$, $|f\rangle$, and $|g\rangle$ in Eq. (\ref{2}) ($P_e$, $P_f$, and $P_g$) for the initial state $|f\rangle$ ($\theta=0$). In Fig. \ref{fl2b1}(a), the populations of the states $|f\rangle$ and $|e\rangle$ decrease from one to zero and the population of the state $|g\rangle$ increases from zero to one. That is because the state $|f\rangle$ is transferred into the state $|g\rangle$ through the transition path $|f\rangle\rightarrow|e\rangle\rightarrow|g\rangle$ due to the dissipation of the cavities. From Eq. (\ref{2}), we know that the atom and the cavity field are in the entangled state, when $0<P_g(t)<1$. In other words, the atom-field system will disentangle, when $P_g(t)=1$. Fig. \ref{fl2b1}(b) shows the evolution of the populations of state $|g\rangle$ over the scaled time $\tau=\kappa t$ for different detunings. It shows that the population of the state $|g\rangle$ increases from zero to one more slowly in the presence of detunings, i.e., the presence of detunings can preserve the entanglement between the atom and the cavity field. It is because that the presence of the detuning $\Delta$ ($\Delta_{l}$) decreases the transition rate between the states $|e\rangle$ and $|g\rangle$ ($|f\rangle$ and $|e\rangle$). As a result, the decay of the entanglement between the atom and the cavity field becomes slow in the presence of detunings.
In addition, the amplitude of the oscillations is associated with the intensity of the memory effect of non-Markovian environments. In Fig. \ref{fl2}, the linear entropy shows more intensive oscillations in the absence of the detuning. In other words, the detunings suppress the memory effect of non-Markovian environments. Hence, the detunings not only make the evolution of the system slow, but also suppress the the memory effect in non-Markovian environments.

\begin{figure}[htbp]
  \centering
\includegraphics[width=0.5\textwidth]{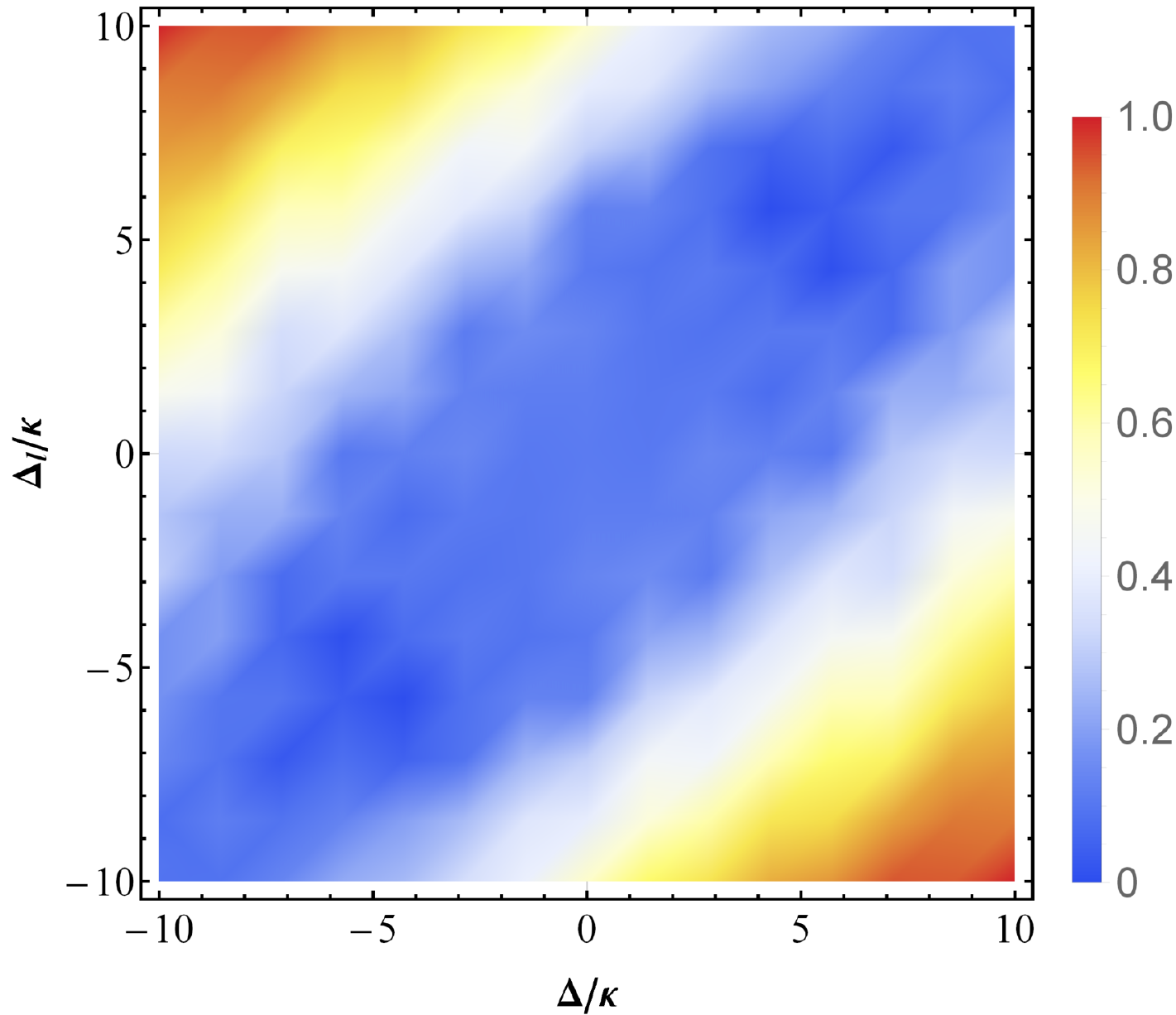}
 \caption{ The linear entropy of the atom-field at the scaled time $\tau=15\kappa t$,
 as a function of detunings $\Delta$ and $\Delta_{l}$ for initial atomic state $|f\rangle$.
 Other parameters: $g=\Omega=10\kappa$.\label{fl2c}}
\end{figure}

In Fig. \ref{fl2c}, we plot the linear entropy of
the atom-field at the scaled time $\tau=15\kappa t$, as a function of the detunings $\Delta$ and
$\Delta_{l}$ for initial atomic state $|f\rangle$. Fig. \ref{fl2c} shows that when the
detuning $\Delta=0$ ($\Delta_{l}=0$), the sign of the detuning $\Delta_{l}$ ($\Delta$) has no
effect on the decay of the entanglement. However, the decay of the entanglement can be
suppressed, when the detunings satisfy the condition $\Delta\cdot\Delta_{l}<0 $. That means
the decay of the entanglement can be suppressed greatly by choosing the sign of the detunings $\Delta$ and $\Delta_{l}$.

\begin{figure}[htbp]
\centering
\includegraphics[width=0.5\textwidth]{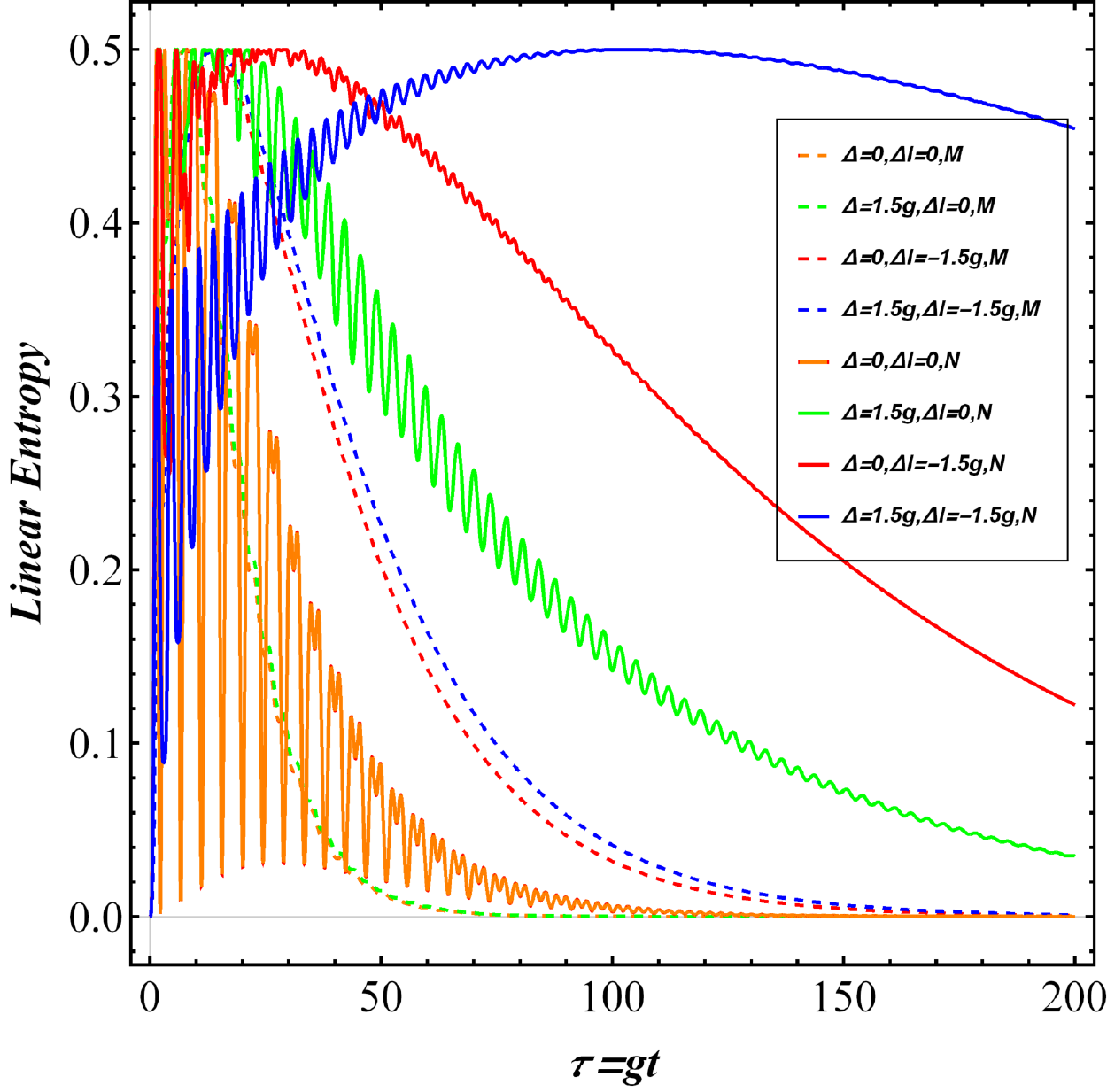}
\caption{ The evolution of the linear entropy between the atom and the cavity field over the scaled time $\tau=gt$ for different detunings in Markovian and non-Markovian environments:
$\Delta=\Delta_{l}=0$ (orange curve),
$\Delta=1.5g$, $\Delta_{l}=0$ (green curve),
$\Delta=0$, $\Delta_{l}=-1.5g$ (red curve), and
$\Delta=1.5g$, $\Delta_{l}=-1.5g$ (blue curve).
The dashed and solid lines denote Markovian and non-Markovian environments, respectively.
Other parameters: $\theta=\varphi=0$.\label{fl2d1}}
\end{figure}

In order to investigate the differences between the Markovian dynamics and the non-Markovian dynamics, we plot the evolution of the linear entropy between the atom and the cavity field over the scaled time $\tau=gt$ in Markovian and non-Markovian environments in Fig. \ref{fl2d1}. It shows that the linear entropy has an obvious oscillation and evolves more slowly in non-Markovian environments for the memory effect. In addition, the presence of detunings can preserve the entanglement between the atom and the cavity field both in the Markovian and non-Markovian environments.

\section{THE ENTANGLEMENT BETWEEN TWO ATOMS}
Due to the fact that the two subsystems are independent, the wave function of the
total system can be written as
\begin{equation}
|\psi(t)\rangle=|\psi_{1}(t)\rangle\otimes|\psi_{2}(t)\rangle.
\end{equation}
However, the atom and the cavity field in each subsystem are in the entangled state. This allows
us to establish the entanglement between the two atoms by performing Bell
state measurement on photons leaving the cavities. We consider the Bell state
\begin{equation}
|\Psi^{+}\rangle=\frac{1}{\sqrt{2}}(|0\rangle_{1}|1\rangle_{2}-|1\rangle_{1}|0\rangle_{2}),
\end{equation}
where $|1\rangle_{i}=\int_{-\infty}^\infty \Theta(\omega)|1_{\omega}\rangle_{i}\mathrm{d}\omega$,
$\Theta(\omega)$ is the pulse shape associated with the coming photon. Then,
acting the projection operator $ P_{F}=|\Psi^{+}\rangle\langle\Psi^{+}| $ on the wave
function $|\psi(t)\rangle$ (after normalization), we obtain
\begin{equation}
\begin{aligned}
|\psi_{AA}(t)\rangle=&\langle\Psi^{+}|\psi(t)\rangle\\
=&\frac{1}{\sqrt{N(t)}}[X_{12}(t)|e,g\rangle-X_{21}(t)|g,e\rangle+Y_{12}(t)|f,g\rangle-Y_{21}(t)|g,f\rangle\\
&+(Z_{12}(t)-Z_{21}(t))|g,g\rangle],
\end{aligned}
\end{equation}
where
\begin{equation}
N(t)=|X_{12}(t)|^{2}+|X_{21}(t)|^{2}+|Y_{12}(t)|^{2}+|Y_{21}(t)|^{2}+|Z_{12}(t)-Z_{21}(t)|^{2}.
\end{equation}
Here we have defined
\begin{equation}
X_{ij}(t)=E_{i}(t)\int_{-\infty}^\infty U_{j}(\omega,t)\Theta^{*}(\omega)\mathrm{d}\omega,
\end{equation}
\begin{equation}
Y_{ij}(t)=F_{i}(t)\int_{-\infty}^\infty U_{j}(\omega,t)\Theta^{*}(\omega)\mathrm{d}\omega,
\end{equation}
\begin{equation}
Z_{ij}(t)=G_{i}(t)\int_{-\infty}^\infty U_{j}(\omega,t)\Theta^{*}(\omega)\mathrm{d}\omega.
\end{equation}
In order to quantify the amount of entanglement between two atoms, we
introduce the negativity \cite{32}, which is defined as
\begin{equation}
N(\rho(t))=\frac{\|\rho^{T_{A}}\|-1}{2},
\end{equation}
where $\rho^{T_{A}}$ is the partial transpose of $\rho$ and $\|X\|\equiv tr\sqrt{X^{\dag}X}$
is the trace norm.
Similarly, we calculate the average negativity with respect to all possible input pure separable states as

\begin{equation}
N^{av}(\rho(t))=\frac{1}{16\pi^{2}}\int N(\rho(t))\prod_{k=1}^{2}\sin(\theta_{k})\mathrm{d}\theta_{k}\mathrm{d}\varphi_{k}.
\end{equation}

\begin{figure}[htbp]
\centering
\subfigure[]{\includegraphics[width=0.4\textwidth]{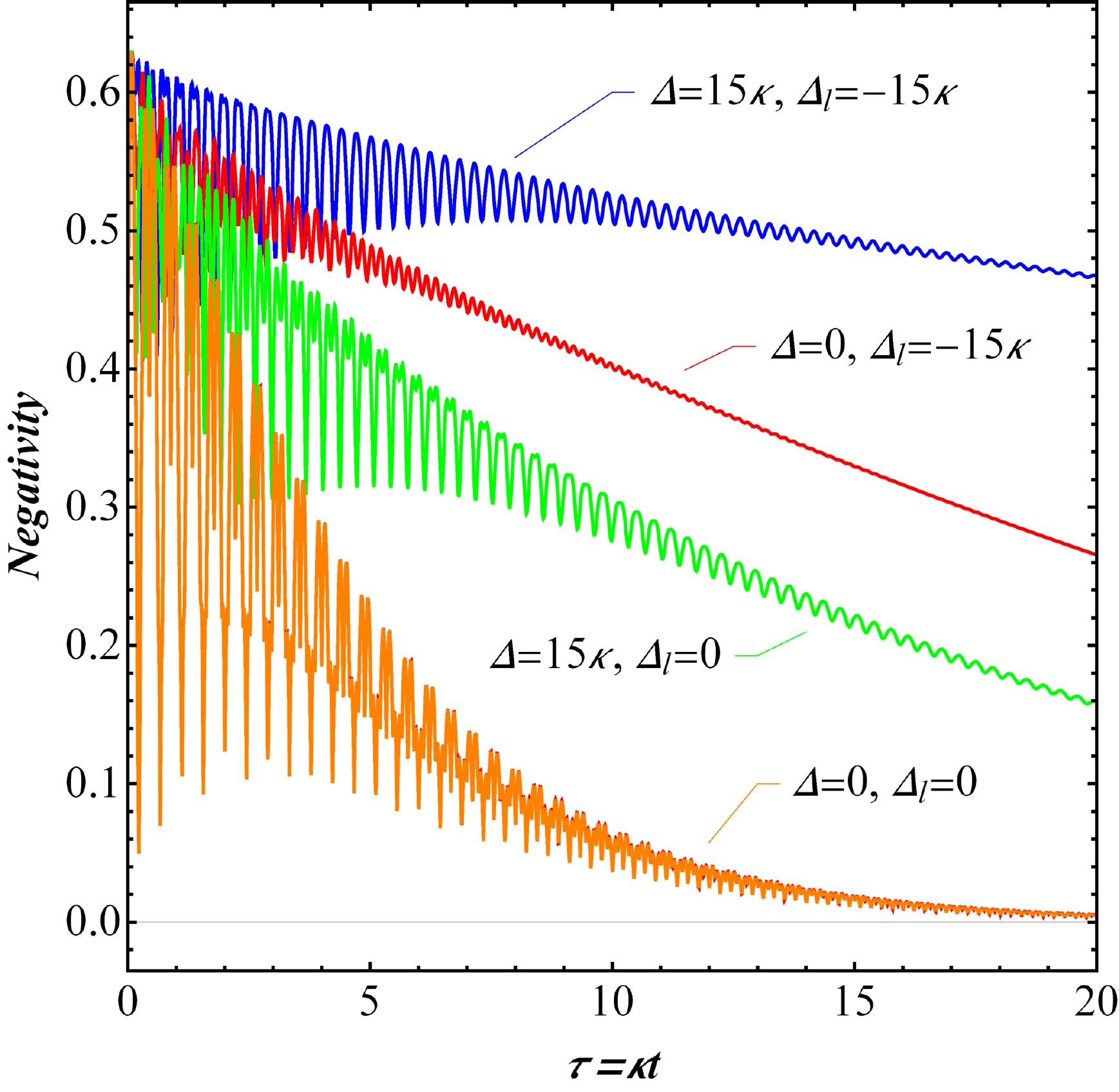}}
\hspace{2mm}
\subfigure[]{\includegraphics[width=0.5\textwidth]{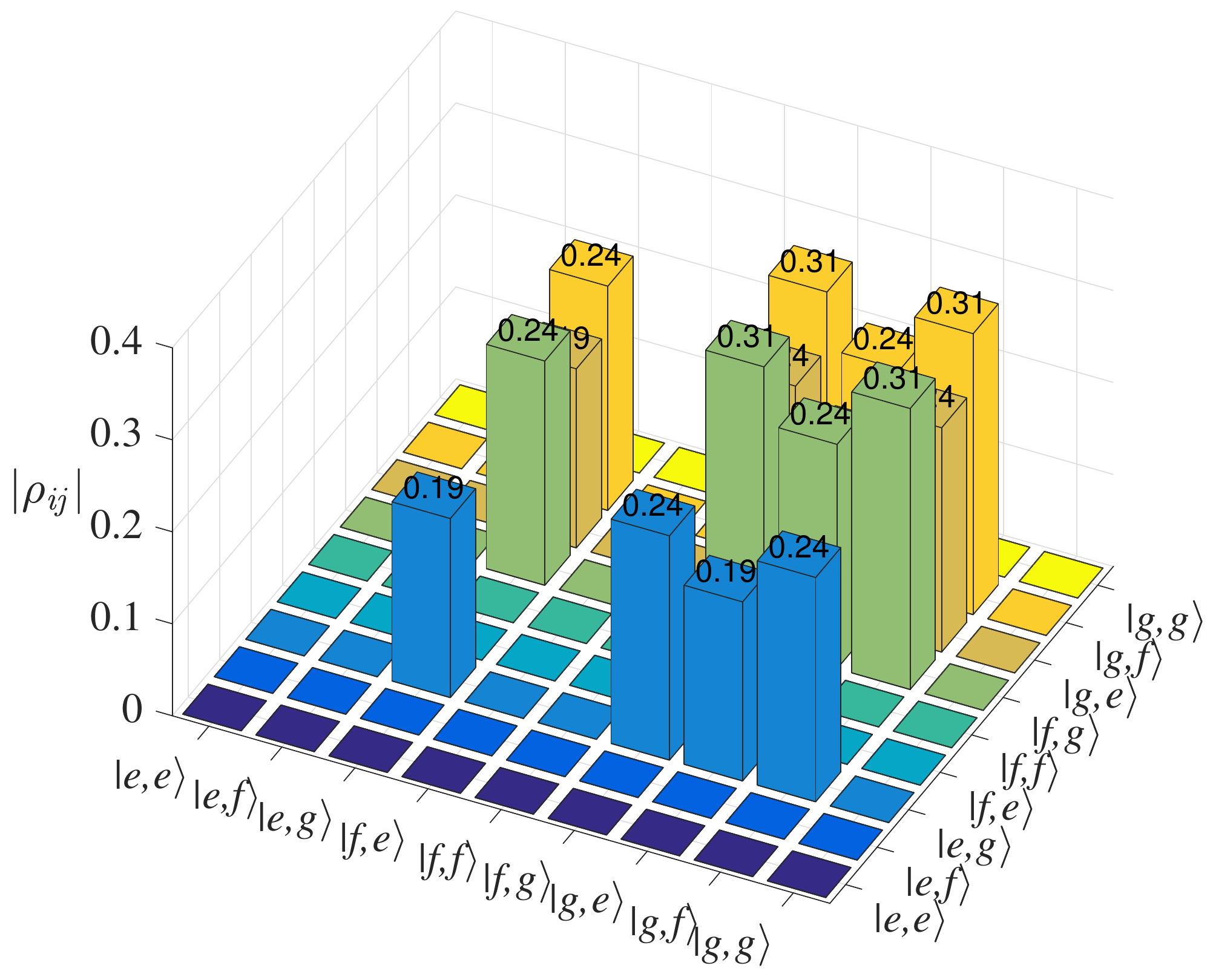}}
\caption{(a) The evolution of the average negativity as a function of the
 scaled time $ \tau=\kappa t $ for different detunings in non-Markovian environments:
$\Delta=\Delta_{l}=0$ (orange curve),
$\Delta=15\kappa,\Delta_{l}=0$ (green curve),
$\Delta=0, \Delta_{l}=-15\kappa$ (red curve), and
$\Delta=15\kappa, \Delta_{l}=-15\kappa$ (blue curve).
(b) Density matrix of the two atoms at $\tau=1\kappa t$:
$\Delta_{1}=\Delta_{2}=15\kappa$,
$\Delta_{l_{1}}=\Delta_{l_{2}}=-15\kappa$,
and $\theta_{1}=\theta_{2}=\varphi_{1}=\varphi_{2}=0$.
Other common parameters: $g=\Omega=10\kappa$.\label{fl3}}
\end{figure}

\begin{figure}[htbp]
\centering
\subfigure[]{\includegraphics[width=0.4\textwidth]{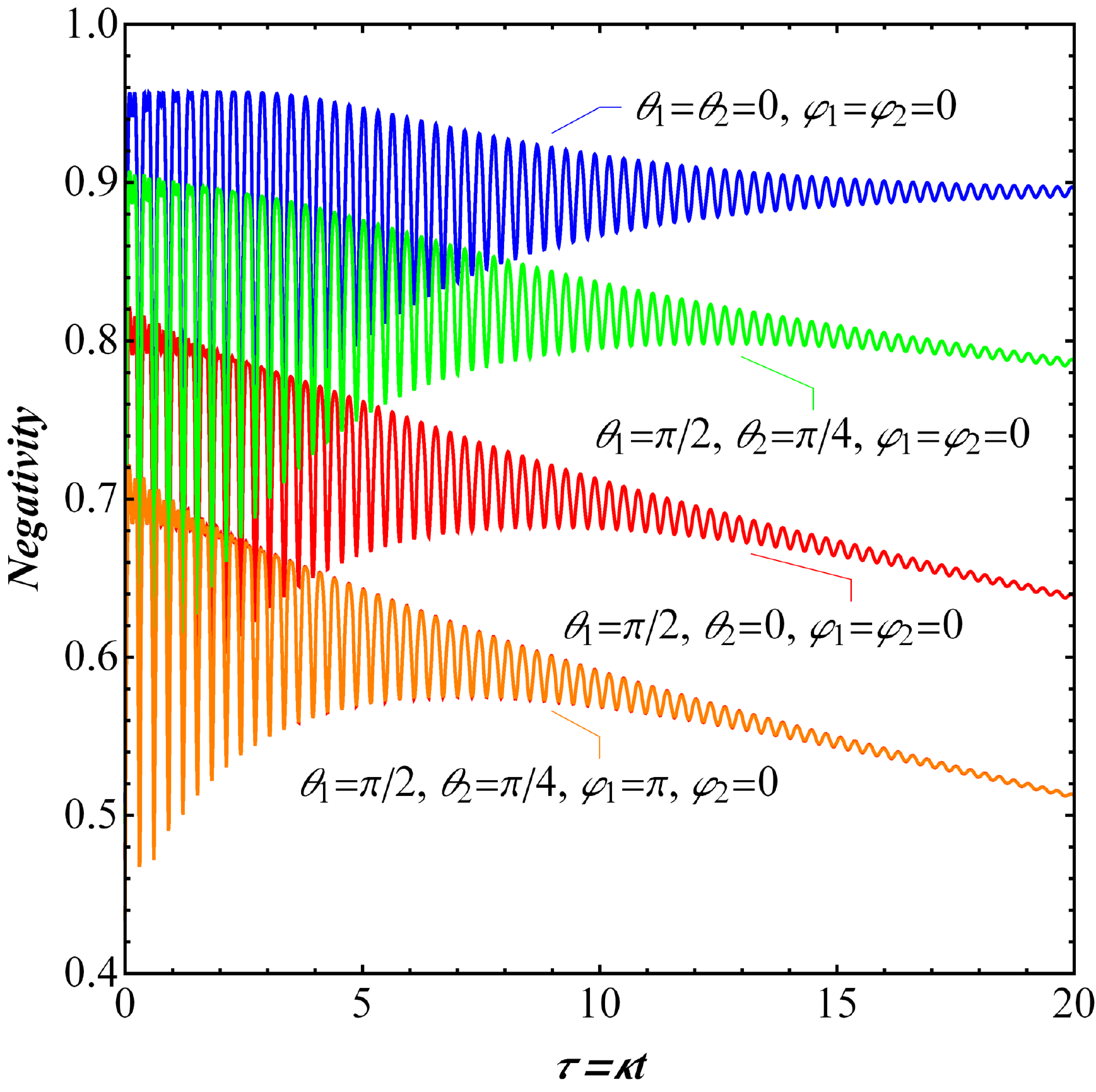}}
\hspace{2mm}
\subfigure[]{\includegraphics[width=0.4\textwidth]{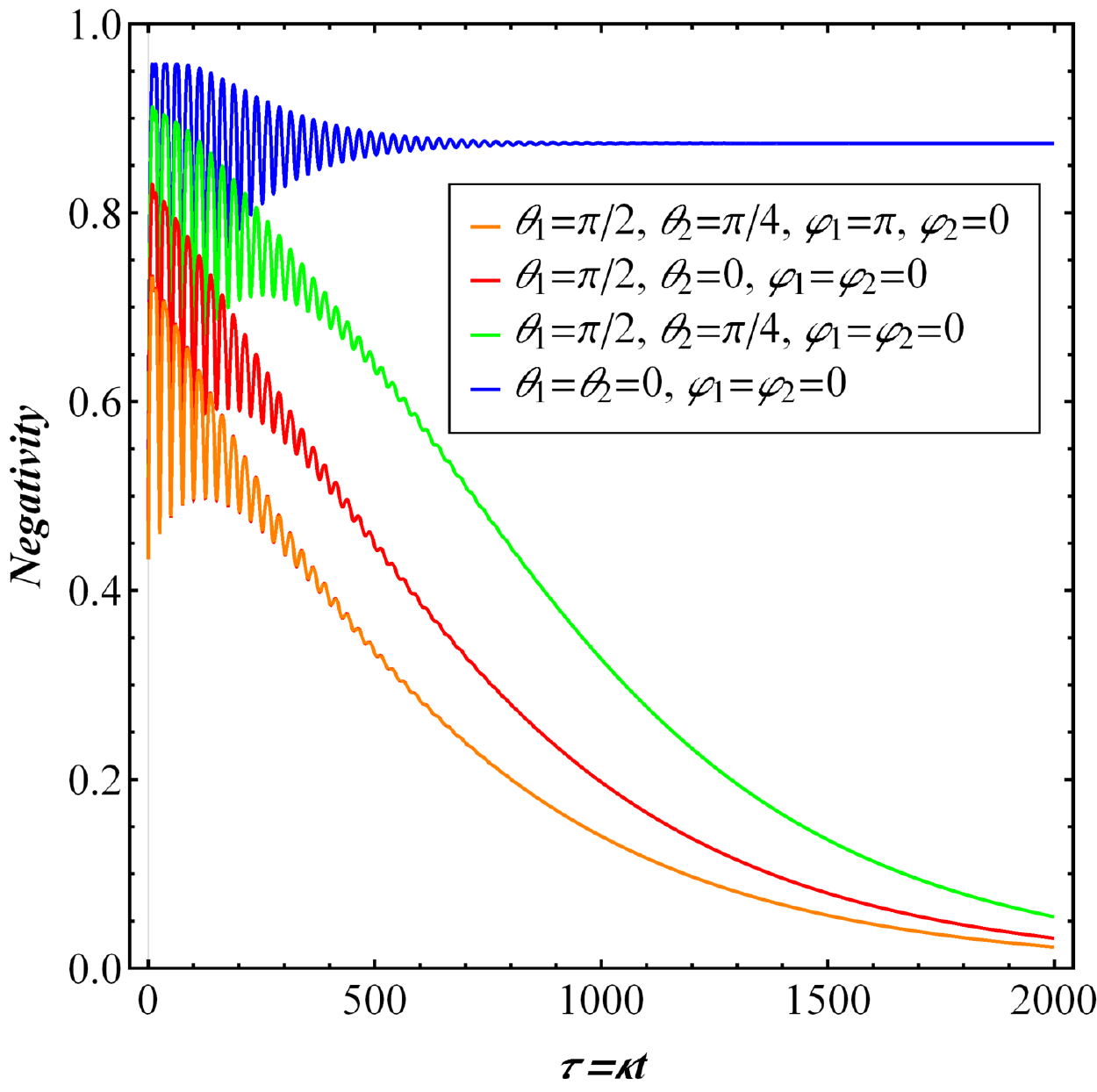}}
\caption{ The evolution of the negativity between the two atoms as a function of the scaled time
$ \tau=\kappa t $ for different initial atomic states in (a) non-Markovian environments and (b) Markovian environments:
$\theta_{1}=\theta_{2}=0$, $\varphi_{1}=\varphi_{2}=0$ (blue curve),
$\theta_{1}=\pi/2$, $\theta_{2}=0$, $\varphi_{1}=\varphi_{2}=0$ (red curve),
$\theta_{1}=\pi/2$, $\theta_{2}=\pi/4$, $\varphi_{1}=\varphi_{2}=0$ (green curve), and
$\theta_{1}=\pi/2$, $\theta_{2}=\pi/4$, $\varphi_{1}=\pi$, $\varphi_{2}=0$ (orange curve).
Other parameters: (a) $\Delta=15\kappa$, $\Delta_{l}=-15\kappa$ and $ g=\Omega=10\kappa $.
(b) $\Delta=0.15\kappa$, $\Delta_{l}=-0.15\kappa$ and $ g=\Omega=0.1\kappa $.\label{fl4}}
\end{figure}

Fig. \ref{fl3} (a) illustrates the evolution of the average negativity, as a
function of the scaled time $\tau=\kappa t$ in non-Markovian environments. In
Fig. \ref{fl3} (a), the average negativity exhibits an oscillatory decay behaviour
in the absence and presence of detuning for the interaction between cavities and
environments. In the presence of the detunings $\Delta$ and $\Delta_{l}$,
the decay of the entanglement between two atoms becomes slow.
It is because that the entanglement between the two atoms depends on the entanglement between the atom and cavity field, i.e., the disentanglement between the atom and the cavity field will lead to the disentanglement between the two atoms. From the results in section 3, we know that the presence of detunings can preserve the entanglement between the atom and the cavity field, namely, the detunings can preserve the entanglement between the two atoms.
Therefore, by
choosing the detunings $\Delta$ and $\Delta_{l}$, a long-living stationary entangled
state between two atoms can be created. Fig. \ref{fl3} (b) shows the density matrix
of two atoms at the scaled time $\tau=1\kappa$.
Due to the dissipation of cavities, the population of the state $|g,g\rangle$ is increased and
the populations of the other states are decreased.

On the other hand, we consider the effect of the initial atomic states on the evolution of the entanglement between two atoms. In Fig. \ref{fl4}, we plot the evolution of the negativity, as a function of the scaled time $\tau=\kappa t$ for different initial atomic states in (a) non-Markovian environments and (b) Markovian environments. It is revealed that, when two atoms are in the quantum
state $|f\rangle$ initially, the long-living stationary entanglement between two atoms is
generated in both Markovian and non-Markovian environments. Furthermore, our further calculations show that when two atoms are in the same quantum state initially, a long-living stationary entangle state between two atoms can be produced in both Markovian and non-Markovian environments.

\section{CONCLUSION}
In summary, we have investigated the system formed by two independent
dissipative cavities, each of which contains a $\Lambda$-type three-level atom.
We solve the time-dependent Sch\"{o}rdinger equation of the subsystem
and obtain the analytical results for the dynamical evolution of the
atom-field system in non-Markovian environments. The results show that
the atom and the cavity field are in the entangled state in non-Markovian
environments and the decay of the entanglement is suppressed
in the presence of detunings. We establish the entanglement between two
atoms by performing Bell state measurement on
photons leaving the cavities. It is revealed that, the presence of
the detunings $\Delta$ and $\Delta_{l}$ can suppress the decay of
the entanglement. By choosing the detunings
and the initial atomic states, a long-living stationary entangled state between two
distant atoms can be generated. Our results are useful to perform
long-distance quantum communication, especially when long-living stationary entanglement is needed
and the effect of environments cannot be neglected.

In the end, we briefly address the feasibility of experimental realization. In our proposed scheme, the two atoms are trapped in two distant cavities, respectively. That can be implemented experimentally by trapping atoms in cavity QED system \cite{37,38}. The leaking photons from the two cavities are mixed on a beam splitter, and two detectors D1 and D2 are set to the both output ports of it. The projection on the states is realized, when one of the two detectors is clicked. The Bell state $|\Psi^{\pm}\rangle=\frac{1}{\sqrt{2}}(|0\rangle_{1}|1\rangle_{2}\pm|1\rangle_{1}|0\rangle_{2})$ can be distinguished by the detectors D1 and D2. The detection scheme can be implemented in linear optical system experimentally \cite{39}.

\section*{Funding}
National Natural Science Foundation of China (NSFC) (11474077, and 11675046), Program for Innovation Research of Science in Harbin Institute of Technology (A201411, and A201412), the Fundamental Research Funds for the Central
Universities (AUGA5710056414), Natural Science Foundation of Heilongjiang Province of
China. (A201303), and Postdoctoral Scientific Research Developmental Fund of Heilongjiang
Province (LBH-Q15060).

\end{document}